\documentclass[9pt, conference]{IEEEtran}
\usepackage{times}
\usepackage{graphicx}
\usepackage{amsmath}
\usepackage{amsfonts}
\usepackage[font=scriptsize]{caption}
\usepackage{subcaption}
\usepackage{float}

\usepackage[numbers]{natbib}
\usepackage{multicol}
\usepackage[bookmarks=true]{hyperref}
\usepackage[colorinlistoftodos]{todonotes}
\begin{document}

\title{Multi-Adversarial Safety Analysis for Autonomous Vehicles}

\author{\IEEEauthorblockN{Gilbert Bahati}
\IEEEauthorblockA{\textit{Civil and Environmental Engineering } \\
\textit{University of California - Berkeley}\\
bahati95@berkeley.edu}
\and
\IEEEauthorblockN{Marsalis Gibson}
\IEEEauthorblockA{\textit{Electrical Engineering and Computer Science} \\
\textit{University of California - Berkeley}\\
mtgibson@berkeley.edu}
\and
\IEEEauthorblockN{Alexandre Bayen}
\IEEEauthorblockA{\textit{Institute of Transportation Studies} \\
\textit{University of California - Berkeley}\\
director-its@berkeley.edu}
}




%

\maketitle
\begin{abstract}
This work in progress considers reachability-based safety analysis in the domain of autonomous driving in multi-agent systems. We formulate the safety problem for a car following scenario as a differential game and study how different modelling strategies yield very different behaviors regardless of the validity of the strategies in other scenarios. Given the nature of real-life driving scenarios, we propose a modeling strategy in our formulation that accounts for subtle interactions between agents, and compare its Hamiltonian results to other baselines. Our formulation encourages reduction of conservativeness in Hamilton-Jacobi safety  analysis to provide better safety guarantees during navigation. 
\end{abstract}

\IEEEpeerreviewmaketitle

\section{Introduction}





\begin{figure}[H]
    \begin{subfigure}[t]{1.5in}
    
        \includegraphics[angle=-90,width=1.3in]{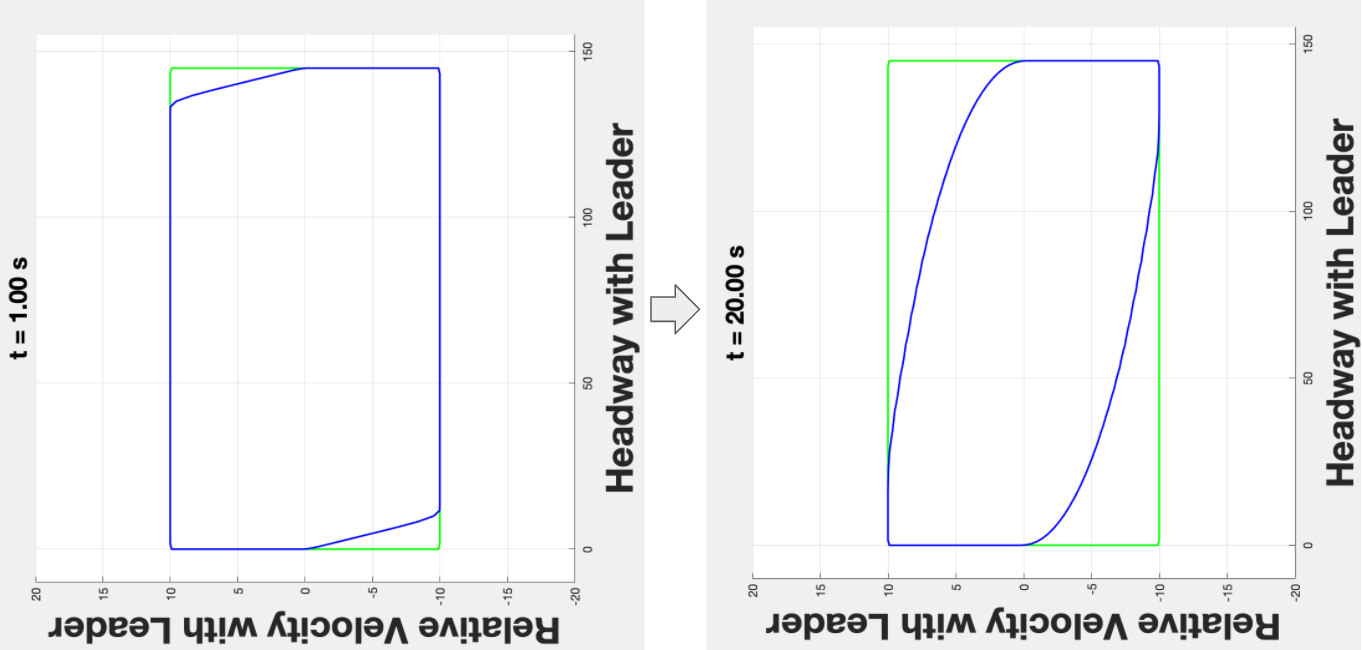}
            \subcaption{Two player scenario (2D state system) in fig \ref{subfig:2carsim}}
            \label{subfig:safeset2D}
    \end{subfigure}
   \quad
    \begin{subfigure}[t]{1.5in}
    
        \includegraphics[angle=-90,width=1.26in]{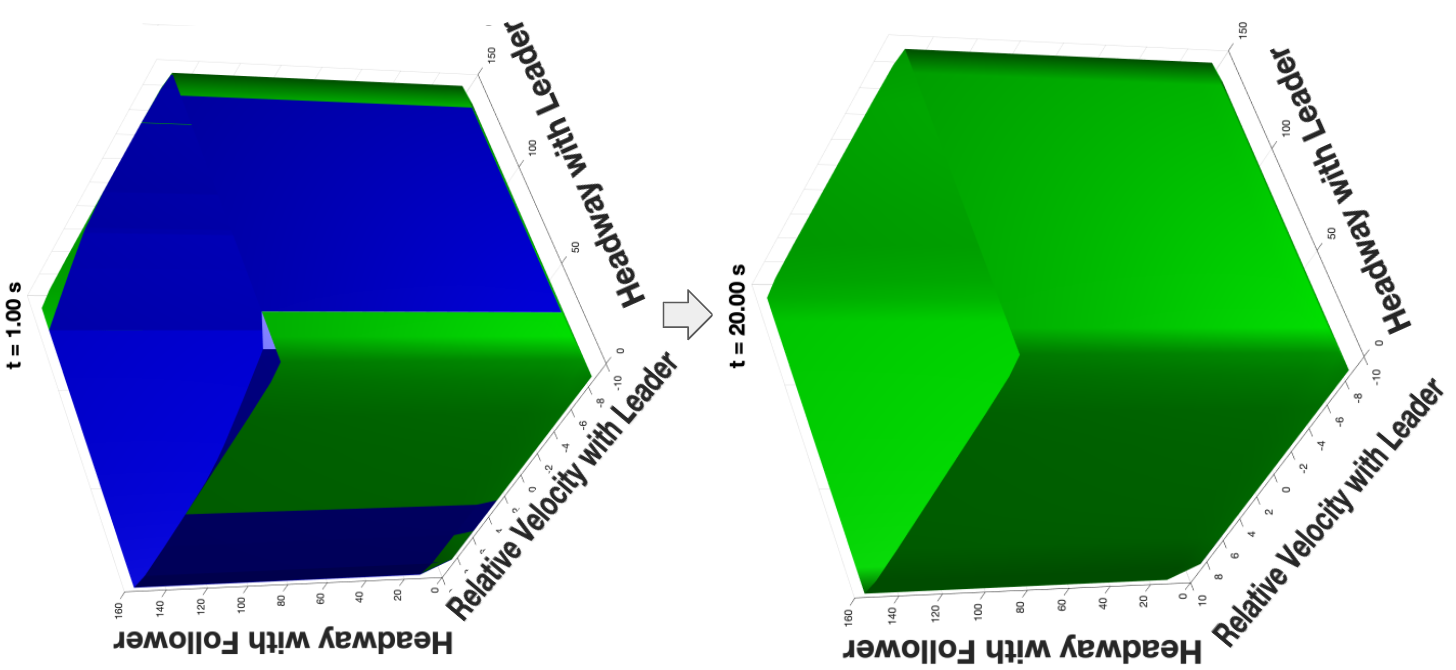}
        \subcaption{Three player scenario (4D state system) in fig \ref{subfig:3carsim}}
        \label{subfig:safeset3D}
    \end{subfigure}
    \caption{HJI reachability analysis \cite{mitchell2005toolbox} computations for fig \ref{fig:illustration}: the constraint set (green) represents our state boundaries and the reachable safe set (blue) remains within/propagates inwards as t increases. To interpret this, a state that starts within the blue safe set is guaranteed to remain within the green constraint set for t seconds. For ease of visualization, \ref{subfig:safeset3D} is a 3D slice of the 4D state system where the relative velocity with follower is 0.}
    \label{fig:sets}
\end{figure}
\begin{figure}[H]
    \begin{subfigure}[t]{1.7in}
    
        \includegraphics[width=1.7in]{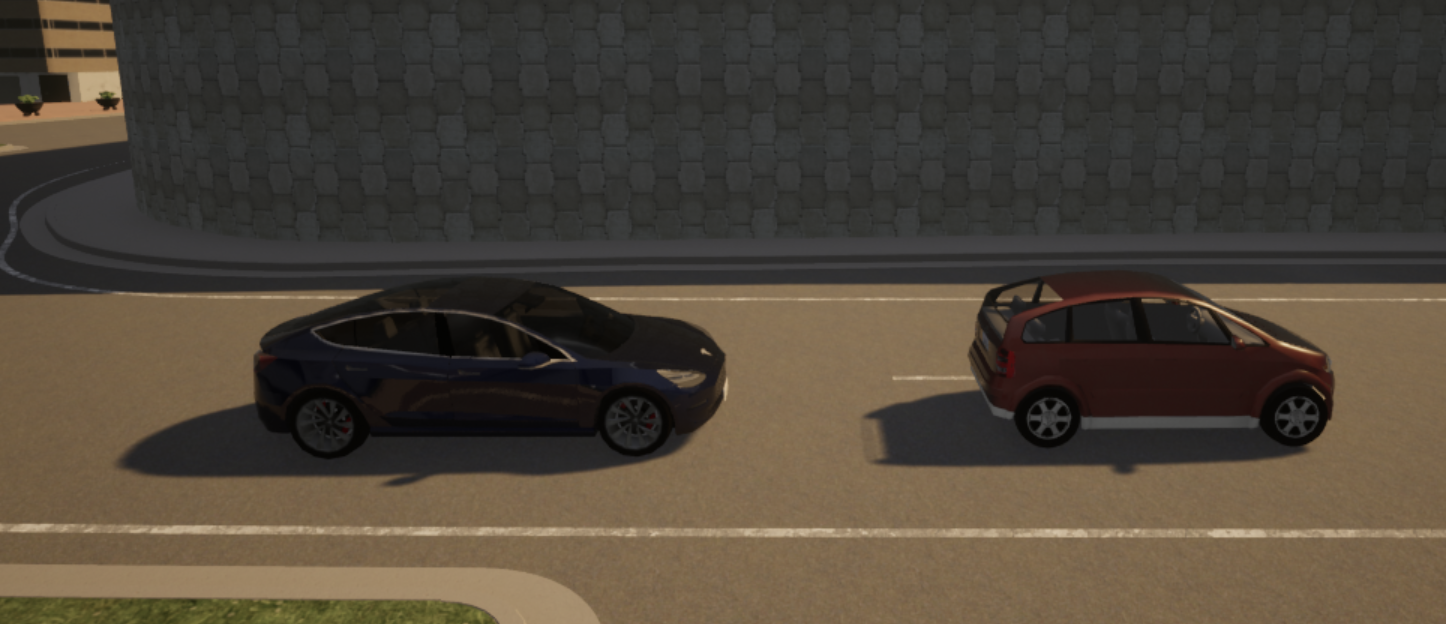}
         \subcaption{Autonomous agent finds optimal strategy}
            \label{subfig:2carsim}
    \end{subfigure}
   \quad
    \begin{subfigure}[t]{1.7in}
    
        \includegraphics[width=1.7in]{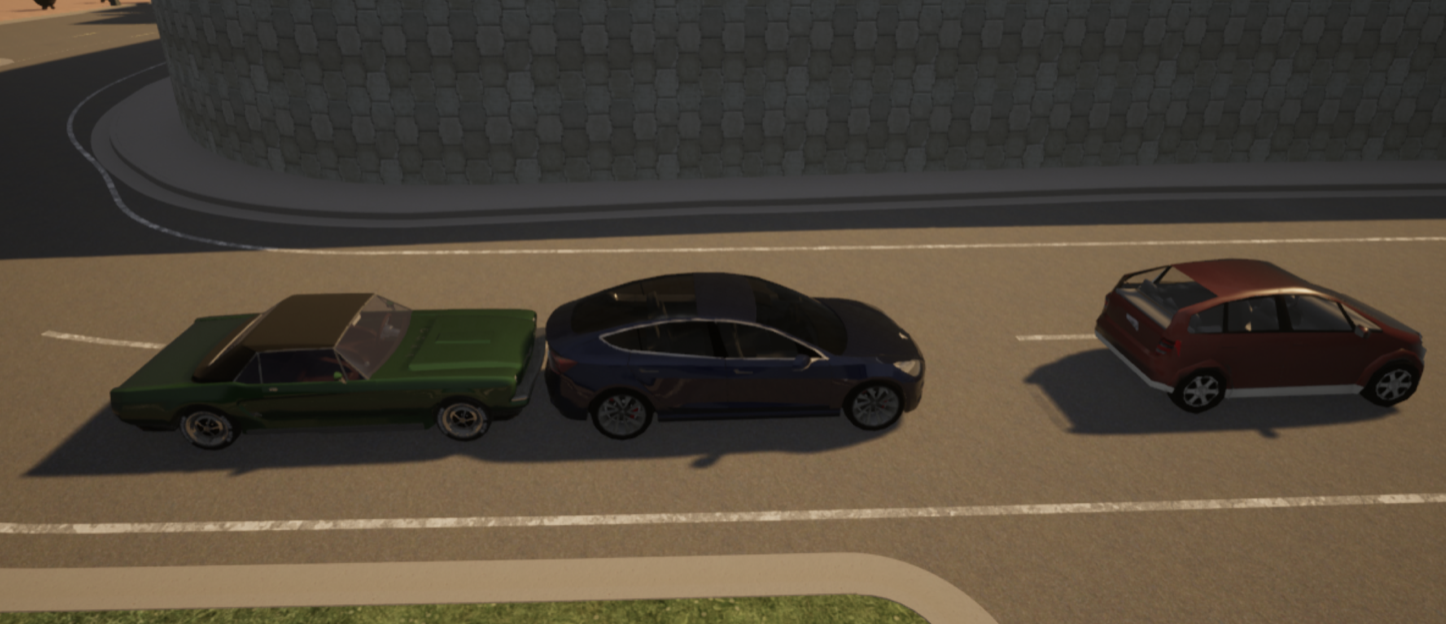}
        \subcaption{Autonomous agent fails to reason and find optimal strategy}
        \label{subfig:3carsim}
    \end{subfigure}

    \caption{The black car represents the autonomous agent while the other two cars represent human drivers. In the two car scenario in \ref{subfig:2carsim}, we compute the safe reachable set under the assumption that the leading car acts adversarially. In this scenario, we obtain a feasible solution, see fig \ref{subfig:safeset2D}. However, this is not the case when a third human driven car is introduced behind the autonomous agent in \ref{subfig:3carsim}. The autonomous agent cannot find a safe reachable set, see fig \ref{subfig:safeset3D}, which means that there are no states or points in time in which the car is guaranteed to be safe if the two human vehicles choose to act adversarially.}
    \label{fig:illustration}
\end{figure}

If autonomous vehicles are to serve as traffic management systems \cite{wu2017flow}, safe navigation around human vehicles on highways and in cities is crucial. However, safe navigation can be difficult to provide because a lot of uncertainty exists in real driving scenarios that complicate the driving problem. Typically,  Hamilton-Jacobi reachability analysis (HJI) can be used to find safe strategies around unknown components of a dynamical system \cite{bansal2017hamilton}. In previous work, researchers develop \cite{fisac2018general} a framework to protect a system against one known source of uncertainty using Hamilton-Jacobi reachability, with the goal of protecting the system from the worst-case scenario. However, in real driving scenarios, it may be necessary to consider multiple sources of uncertainty. As depicted in figure \ref{fig:sets} and \ref{fig:illustration}, extreme worst-case scenarios may never provide a feasible safety strategy, and it may be the case that establishing safety is impossible.

Therefore, in this work, we study the reduction of conservativeness in Hamilton-Jacobi safety analysis by introducing structure into some or all of the human models. Specifically, we study a modeling strategy around the second disturbance that takes advantage of the structure of human behavior in a way that allows us to use differential game theory in more dense dynamic driving environments.

\section{System dynamics}

\begin{figure}[H]
\includegraphics[scale=0.15]{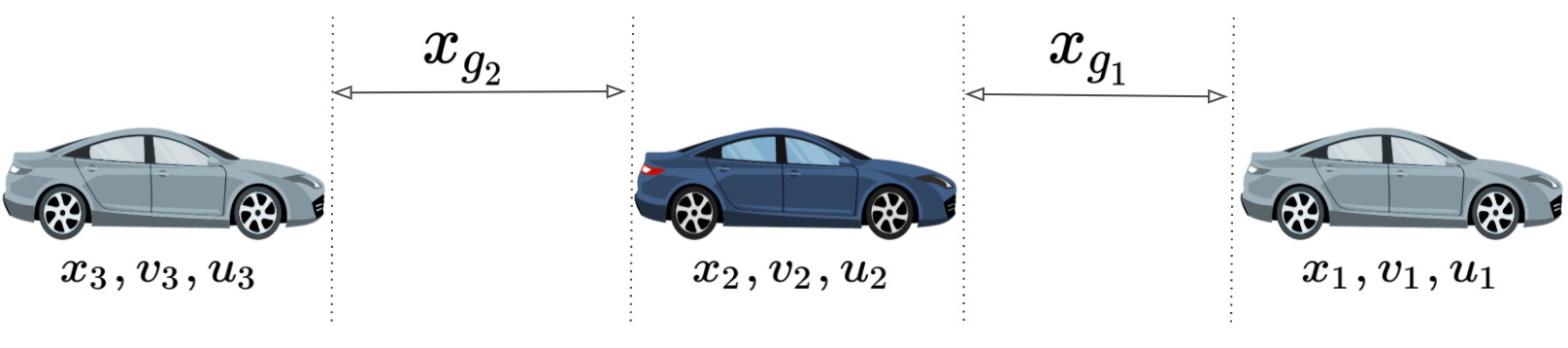}
\caption{Car following scenario where the subject vehicle drives in between two human vehicles.}
     \label{fig:scenario}
\end{figure}
We consider a dynamical system with state $z \in \mathbb{R}^n$, and three inputs, $u \in \mathcal{U} \subset \mathbb{R}^{n_u}$, $d_1 \in \mathcal{D}_1 \subset \mathbb{R}^{n_d}$, $d_2 \in \mathcal{D}_2 \subset \mathbb{R}^{n_d}$, which we refer to as the controls, disturbance 1, and disturbance 2 respectively. Our system dynamics are generally defined as: 
\begin{equation}
    \dot{z} = f (z,u,d_1,d_2)
\end{equation}
Disturbance 1 and 2 represent the uncertainty around the leading and following human vehicle respectively. In our car following scenario in figure \ref{fig:scenario}, the goal for the autonomous agent is to establish safety and remain in between the other two players given their actions. Thus, the dynamics between all three vehicles can be described using their relative position $x_j$, relative speed $v_j$, and relative accelerations $u_i$ as in the following: 





\begin{equation}
    \begin{matrix}
    \begin{bmatrix}
        \dot{x}_{g_1} \\ 
        \dot{v}_{g_1}\\
        \dot{x}_{g_2}\\
        \dot{v}_{g_2}\\
    \end{bmatrix} = 
    
    \begin{bmatrix}
        v_{g_1} \\
        u_1 - u_{2}\\
        v_{g_2}\\
        u_{2} - u_3\\
    \end{bmatrix}
    \end{matrix}
\end{equation}

s.t.
\begin{equation*}
u_{i} \in [a_{min}, a_{max}], \ \forall \ i = 1...3 \\
\end{equation*}
\begin{equation*}
x_{j} > 0, \ \forall \ j = g_1, g_2\\
\end{equation*}
                

In the next section, Section III, we discuss how we choose our uncertainty and how we pose our safety problem.

\section{Three-Player Differential Game}
The safety problem is posed as a differential game between three players, where the system controller, $u$, plays against two adversaries, $d_1$ and $d_2$, also known as the system's uncertainty. To obtain a \textit{safe policy} for the system, we chose a function, $l(\textbf{z})$, that assigns a safety value to the current state, $\textbf{z}$ and formulate a game whose outcome is given by the function $\mathcal{V}: \mathbb{R}^n \times \mathcal{U} \times \mathcal{D}_1 \times \mathcal{D}_2 \rightarrow \mathbb{R}$. $\mathcal{V}$ assigns each initial state $\textbf{z}$ and player strategies $u(\cdot)$, $d_1(\cdot)$, $d_2(\cdot)$, the lowest value of $l(\cdot)$ ever achieved by a trajectory $\xi^{\textbf{z}, u}_{d_1, d_2}(\cdot)$ from state $\textbf{z}$.
\begin{equation}
    \mathcal{V}(\textbf{z}, u(\cdot), d_1(\cdot), d_2(\cdot)) = \inf_{t\geq 0} l(\xi^{\textbf{z}, u}_{d_1, d_2}(t))
\end{equation}

The goal of system is to maximize the objective, while the goal of the active adversaries is to minimize the objective. Thus, the game formulation that we want to solve is\footnote{Technically, as in \cite{mitchell2005toolbox}, we restrict each disturbance to a set of nonanticipative strategies. Therefore, $d_1(\cdot)$ and $d_2(\cdot)$ in eq. \ref{eq:valuefunc} are actually maps, $\beta_1[u(\cdot)](\cdot)$ and $\beta_2[u(\cdot)](\cdot)$, that respectively maps our control input to their corresponding disturbance input.}:
\begin{equation}
    \label{eq:valuefunc}
    V(\textbf{z}) = \inf_{d_1, d_2} \sup_{u} \mathcal{V}(z, u(\cdot), d_1(\cdot), d_2(\cdot)) \\
\end{equation}

\subsection{Player Strategies}


We formulate uncertainties $d_1$ and $d_2$ around the two human driving actions, for example $u_1$ and $u_3$, to represent behavioral properties that are trying to perturb the autonomous system. More specifically:

\begin{enumerate}
    \item First, we consider a \textbf{baseline} assignment: 
    \begin{itemize}
        \item $d_1 = u_1$ where $u_1 \in [a_{min}, a_{max}]$
        \item $d_2 = u_3$ where $u_3 \in [a_{min}, a_{max}]$
    \end{itemize}
    \item Then, we consider an \textbf{alternative} assignment for $d_2$ by taking advantage of the structure of human driving and modeling $u_3$ using a car following model $g(z, d_2)$: 
    \begin{itemize}
        \item $d_1= u_1$ where $u_1 \in [a_{min}, a_{max}]$
        \item $u_3 = g(z, d_2)$
    \end{itemize}
\end{enumerate}

In our second strategy, $d_2$ uses psycho-physiological characteristics in human driving as an alternative modeling strategy \cite{TrafficDynamics}. Additionally, we ensure that the values of $u_{3}$ are within realistic bounds given all possible autonomous agent's actions $u_{2}$. This modelling strategy relaxes unrealistic extremities of the previous dynamic game formulation and implicitly models interaction effects between agents for realistic safety. We model the following vehicle's driving behavior using the Intelligent Driver's car following model and explicitly model $d_2$ as \textbf{safe-reaction time, \textit{T}}, as follows: 
\begin{equation}
g(z,T) = a \bigg( 1 - \bigg( \frac{v_3}{v_0} \bigg) ^\delta - \bigg( \frac{s^*(z,T )}{x_{g_2}} \bigg)^2 \bigg)
\end{equation}
\begin{equation}
s^*(z, T)  = s_0 + \max \bigg( 0, v_3 T + \frac{v_3 (-v_{g_2})}{2 \sqrt{ab}}  \bigg)
\end{equation}
where: \\
$T$: safe reaction-time $(ie.\ d_2 = T\ \in \ [0, T_{max}])$    \\
$s^*(z, T)$: desired headway of the following vehicle \\
$s_0$: minimum desired headway (ie. $s_0 = 0$ to allow crashes)\\
$a, b$: maximum acceleration and deceleration respectively\\
$\delta, v_0$: acceleration exponent (usually 4) and desired velocity respectively

\subsection{Resulting policies}
The optimal control strategy, $u_2^*$, and the optimal disturbance strategy for the first human vehicle, $ d_1^*$ are calculated from (4) using the Hamiltonian numerics to be:
\begin{align}
u_2^* = 
\begin{cases}
    a_{max} &\mbox{if } (p_4 - p_2) > 0 \\
    a_{min} &\mbox{else }
\end{cases}
\end{align}
\begin{align}
d_1^* = 
\begin{cases}
    a_{max} &\mbox{if } p_2 < 0 \\
    a_{min} &\mbox{else }
\end{cases}
\end{align}
and the optimal disturbance strategy for the second human vehicle  is likewise calculated to be: 
\begin{align}
\text{Baseline } 
d_2^* = 
\begin{cases}
    a_{max} &\mbox{if } p_4 > 0 \\
    a_{min} &\mbox{else }
\end{cases}
\end{align}

\begin{align}
\text{Alternative }
d_2^* = 
\begin{cases}
    \min(T_{max}, \max(T_{min}, \frac{-v_{g_2}}{2\sqrt{ab}})) &\mbox{if } p_4 > 0 \\
    \displaystyle\max_{T}| 2\sqrt{ab}T+v_{g_2} | &\mbox{else }
\end{cases}
\end{align}
where: $p_2$ = $\nabla_{v_{g_1}}V$ and $p_4$ = $\nabla_{v_{g_2}}V$. 

\section{Results and Conclusion}
\begin{figure}[H]
    \begin{subfigure}[t]{1.7in}
    
        \includegraphics[angle=-90,width=1.65in]{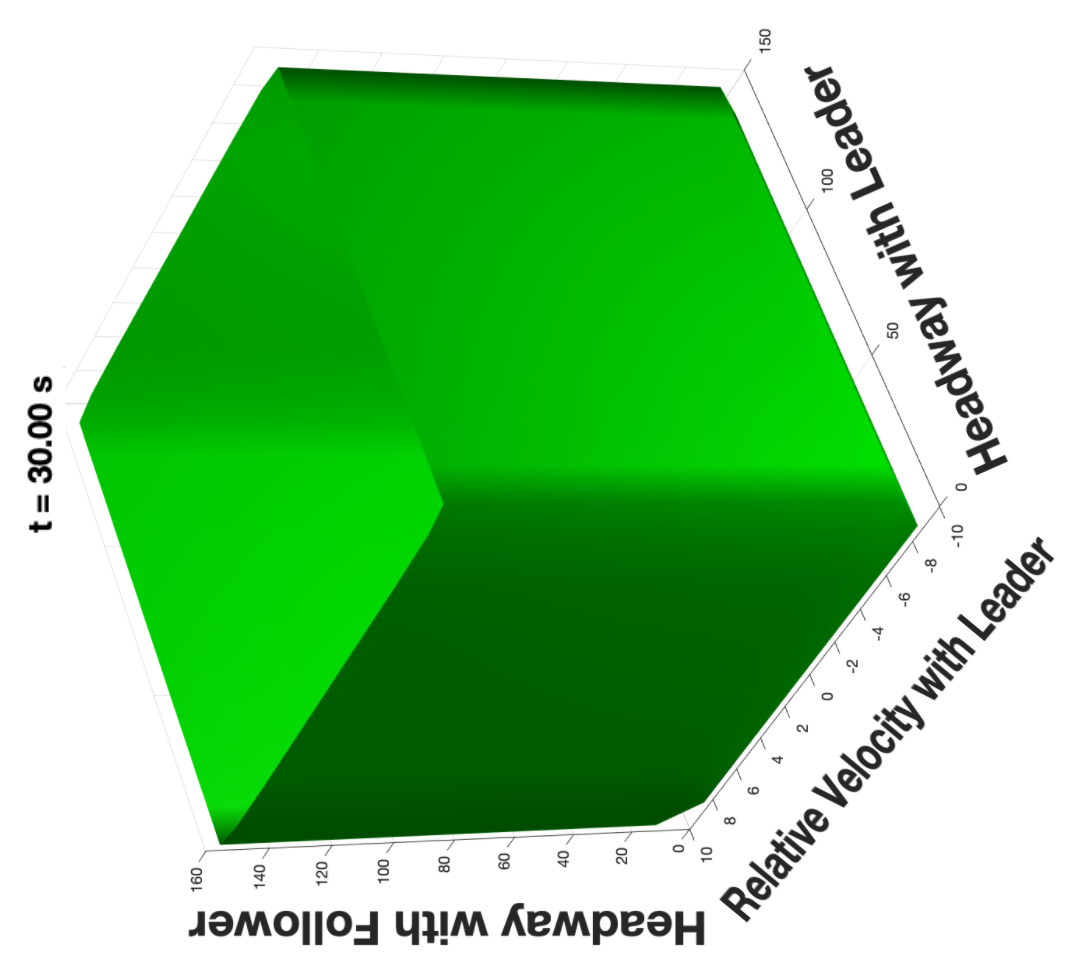}
            \subcaption{ $d_2$ = Extreme Actions (baseline)}
    \end{subfigure}
   \quad
    \begin{subfigure}[t]{1.5in}
    
        \includegraphics[angle=-90,width=1.6in]{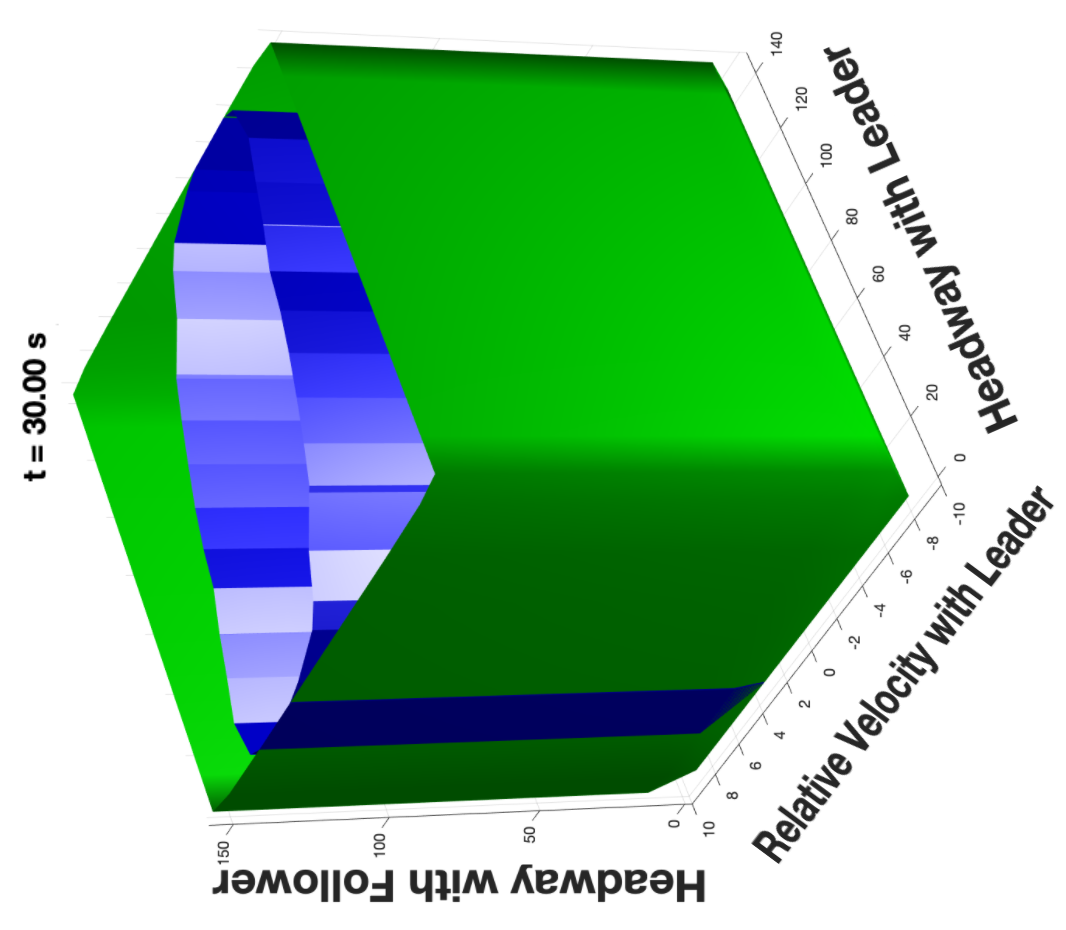}
        \subcaption{$d_2$ = Reaction Time}
    \end{subfigure}
    \caption{Invariant safe states resulting from the baseline and alternative technique.}
    \label{fig:safe sets}
    \vspace{-1em}
\end{figure}

As depicted in figure \ref{fig:safe sets}, the alternative formulation for the second disturbance is able to uncover hidden safe strategies for the 3-car scenario. Setting the disturbance and control bounds to [-1.5, 1.5] and [-2, 2] respectively, the baseline produces an empty blue set, meaning there are no guaranteed safe states for the simulation with this particular disturbance/control setting. However, by taking advantage of psycho-physiological characteristics and driver influences on the road, we discover that a non-empty blue set does exist.
In conclusion, when considering worst-case uncertainty in human driving, if we are to guarantee safety in the chaotic world of driving, we may need to incorporate better information structures of human behavior in our analysis and update our assumptions as we uncover more knowledge about the system. However, by choosing a specific model structure to reduce conservativeness of reachability analysis, we run the risk of not being able to capture human behavior some of the time  due to the limitations of our chosen model ( which serves to capture only approximations). The performance of a chosen model will vary greatly depending on the particular type of driver and circumstance. Therefore, to tackle this trade-off, we further aim to incorporate real-time analysis and data-driven models to learn disturbances (and how they accurately evolve), and maintain formal and robust safety guarantees using different learning strategies such as deep reinforcement learning.

\section*{Acknowledgments}
This material is based upon work supported by the U.S. Department of Energy’s Office of Energy Efficiency and Renewable Energy (EERE) under the Vehicle Technologies Office award number CID DE-EE0008872. The views expressed herein do not necessarily represent the views of the U.S. Department of Energy or the United States Government. 


\bibliographystyle{plainnat}
\bibliography{main}

\end{document}